\documentclass[conference]{IEEEtran}

\usepackage[utf8]{inputenc}

\usepackage[cmex10]{amsmath}

\usepackage[capitalise]{cleveref}
\crefformat{equation}{(#2#1#3)}

\ifCLASSOPTIONcompsoc  \usepackage[nocompress]{cite}
\else                  \usepackage{cite}
\fi

\usepackage{url}

\usepackage{graphicx}

\ifCLASSOPTIONcompsoc \usepackage[caption=false,font=normalsize,labelfont=sf,textfont=sf]{subfig}
\else \usepackage[caption=false,font=footnotesize]{subfig}
\fi

\graphicspath{{Figures/}}

\begin{document}

\title{UAV-Assisted Wireless Communications: An Experimental Analysis of A2G and G2A Channels}

\author{
	\IEEEauthorblockN{Kamran Shafafi, Eduardo Nuno Almeida, André Coelho, Helder Fontes, Manuel Ricardo, Rui Campos}
	\IEEEauthorblockA{INESC TEC and Faculdade de Engenharia, Universidade do Porto, Portugal\\
	\{kamran.shafafi, eduardo.n.almeida, andre.f.coelho, helder.m.fontes, manuel.ricardo, rui.l.campos\}@inesctec.pt}
}

\maketitle

\begin{abstract}
Unmanned Aerial Vehicles (UAVs) offer promising potential as communications node carriers, providing on-demand wireless connectivity to users. While existing literature presents various wireless channel models, it often overlooks the impact of UAV heading. This paper provides an experimental characterization of the Air-to-Ground (A2G) and Ground-to-Air (G2A) wireless channels in an open environment with no obstacles nor interference, considering the distance and the UAV heading. We analyze the received signal strength indicator and the TCP throughput between a ground user and a UAV, covering distances between 50~m and 500~m, and considering different UAV headings. Additionally, we characterize the antenna's radiation pattern based on UAV headings. The paper provides valuable perspectives on the capabilities of UAVs in offering on-demand and dynamic wireless connectivity, as well as highlights the significance of considering UAV heading and antenna configurations in real-world scenarios.\looseness=-1
\end{abstract}


\section{Introduction} \label{sec:Introduction}

Due to the decreasing costs, size, and weight, as well as the increasing endurance, high maneuverability, and ability to hover, Unmanned Aerial Vehicles (UAVs) have emerged as interesting platforms for a wide set of applications, such as surveillance, aerial imagery, operations in unreachable areas, delivery of goods, and search and rescue missions~\cite{shafafi2023joint}. A key capability envisioned by 5G and beyond cellular networks is quickly deploying and providing on-demand temporary wireless connectivity in emergencies and crowded areas. In this regard, the use of UAVs forming aerial wireless networks has been noted as a cost-effective and flexible solution to carry network hardware and establish a temporary network infrastructure, providing wireless connectivity and enhancing the capacity of existing networks~\cite{9170768}. Sending non-critical data, such as video transmission,  requires the maximization of the throughput between the users on the ground and the UAV while tolerating errors and delays. However, sending critical data such as control signals (e.g., controlling robots for search and rescue missions, UAV controls) requires high Quality of Service (QoS), namely low delay and Packet Loss Ratio (PLR)~\cite{HUDA2022103341}. In addition to establishing a reliable link, the QoS requirements such as throughput, PLR, and delay should be considered~\cite{PUNDIR2021103084}. Still, establishing reliable broadband wireless links faces many challenges. In this regard, wireless channel modeling and characterization are important for designing and optimizing wireless communications systems. Accurate models allow the prediction of the signal quality, the evaluation of the system performance, and the development of new wireless technologies. They enable network designers to identify potential problems and optimize system parameters, making wireless channel modeling a critical area of research in wireless communications.\looseness=-1

Comprehensive channel modeling and channel characteristic measurements are essential to ensure a reliable broadband wireless connection. Existing research has primarily focused on Air-to-Ground (A2G) channel modeling, and little attention has been given to critical parameters such as antenna orientation, the Effective Radiation Pattern (ERP) of the antenna system considering the influence of the UAV body, receiver altitude, and UAV heading. These parameters significantly impact wireless channel performance in various environments and altitudes. To address this gap, these parameters should be integrated into channel models and conduct real-world experimental measurements to validate their accuracy. Optimizing antenna orientation, considering receiver altitude, and accounting for UAV heading can improve signal strength and communication performance.\looseness=-1

The main contribution of this paper is the experimental characterization of the A2G and Ground-to-Air (G2A) wireless channels in an open environment with no obstacles nor interference. The characterization of the channel is performed at different distances and UAV headings within the context of the H2020 ResponDrone\footnote{https://respondroneproject.com/} project~\cite{9925792}. We analyze the Received Signal Strength Indicator (RSSI) and the TCP throughput between a user on the ground and a UAV for distances ranging from 50~m to 500~m, and considering different headings of the UAV. Moreover, we provide a more accurate characterization of the channel model, when compared to deterministic models, such as Friis and two-ray. Finally, we characterize the ERP of the antennas based on the headings of the UAV.\looseness=-1

The remainder of this paper is organized as follows. The related work is presented in Section II. The system setup is described in Section III. The field trial to model the experimental wireless channels is presented in Section IV. The experimental results are analyzed in Section V. Finally, Section VI draws conclusions and points out future work.\looseness=-1

\section{Related Work} \label{sec:Related Work}
In the context of A2G communications channel characterization for UAVs, different measurement models and approaches are presented in the literature. In the study conducted by~\cite{8787874}, the authors performed a comprehensive analysis of A2G, Ground-to-Ground~(G2G), and Air-to-Air~(A2A) channel measurements and models, specifically focusing on civil aeronautical and UAV communications. Their work primarily delved into the link budget analysis of UAV communications, where they examined and presented design guidelines to effectively manage communication links, taking into account propagation losses and link fading. In~\cite{ALMEIDA2021102525}, the authors presented a model for the characterization of the A2G and G2A channels, while the UAV was hovering at different altitudes including different Line-of-Sight (LoS) distances from the User Equipment (UE). The channel model has been analyzed in terms of path loss and fast-fading components. In~\cite{9170768}, A2G channel measurements are presented for small-sized UAVs at different environmental conditions and altitude values between 15~m and 105~m. Path loss, shadow fading, Doppler effect, Power Delay Profile, Root-Mean-Square (RMS) delay spread, RMS Doppler frequency spread, and the Rician K-factor were used to characterize the channel. While their investigation provided valuable insights into the general characteristics of these wireless channels, they did not extensively explore the impact of UAV heading on channel behavior, which is crucial in dynamic UAV scenarios.\looseness=-1

In~\cite{8411465}, the authors reviewed empirical models for the A2G and A2A propagation channels. Then, they classified the UAV channel modeling approaches as deterministic, stochastic, and geometric–stochastic models. In~\cite{8981931}, the authors present experimental results on how the height of the receiver affects Radio Frequency~(RF) signal propagation over the sea and the capacity of the radio link. In~\cite{7928222}, the authors proposed an architecture of a new model for the A2G channel. The modeling is based on 10~MHz channel-sounding flight measurements. The key advantage of the proposed A2G channel modeling approach is its flexibility to a wide range of potential ground station deployment scenarios. In~\cite{8740242}, experimental results in commercial Long-Term Evolution~(LTE) deployments were conducted to evaluate the variation of the mean Angle of Arrival~(AoA) and Angular Spread~(AS) with flying altitude. The authors of ~\cite{8740242} used sixteen antennas and LTE cellular signals, to experimentally evaluate the performance of the A2G channel, taking into account the UAV altitude variation. Maximum ratio combining and conventional beamforming techniques have been compared with a single antenna system. In~\cite{inproceedings}, an experimental measurement campaign for the A2G channel at 10~MHz is presented at short (30~m~--~330~m) and long (9~km~--11~km) distances between the receiver and the transmitter. In~\cite{6214705}, measurements with a helium balloon in stationary positions at altitudes up to 500~m have been considered for the A2G channel model. The Euclidean distance between the base station and the receiver was 1900~m. The experiments have been conducted using passive sounding of Universal Mobile Telecommunications System (UMTS) and Global System for Mobile Communications (GSM) signals in an urban environment at a central carrier frequency of 2120~MHz. Using RSSI values, in~\cite{yanmaz6162389}, the authors calculated the path loss exponent for A2G networks while the UAV was flying over both an open field and a campus area. The authors also measured the UDP throughput of the Air-Ground-Air~(AGA) links.\looseness=-1

In conclusion, to the best of our knowledge, the existing literature lacks extensive exploration of the influence of UAV heading on wireless channels. To address this limitation, our research aims to provide an experimental characterization of the A2G and G2A wireless channels, considering both distance and UAV heading as critical parameters. By integrating UAV heading as a key parameter in channel characterization, we aim to gain valuable insights into the capabilities of UAVs in offering dynamic wireless connectivity, while also addressing the limitations of previous works in handling UAV heading variations.\looseness=-1

\begin{figure}
	\centering
	\includegraphics[width=\linewidth]{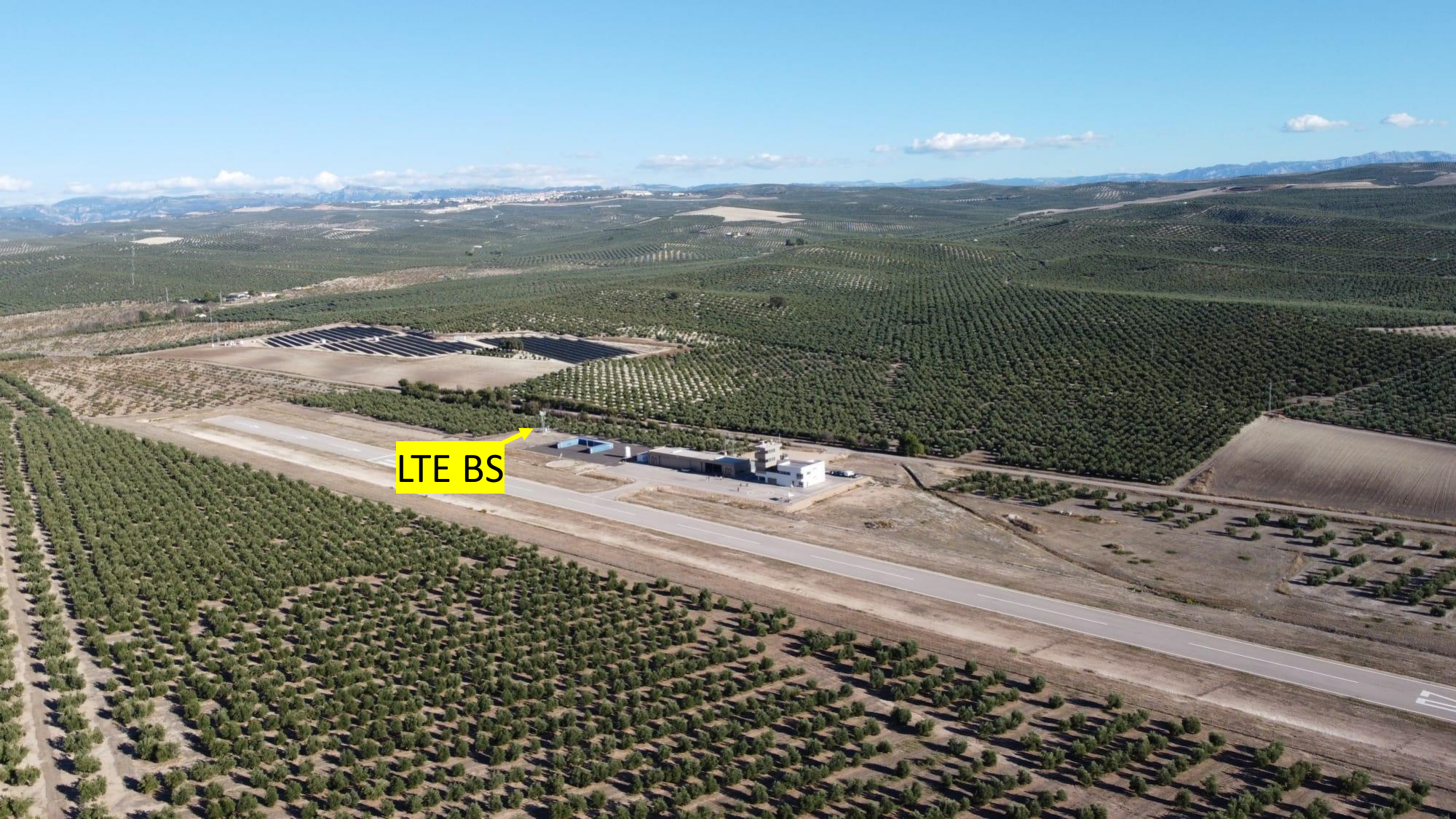}
	\caption{ATLAS site, showing the hangar and runway near which the experimental measurements were performed.}
	\label{fig1:hangar}
\end{figure}

\begin{figure}
    \centering
    \includegraphics[width=\linewidth]{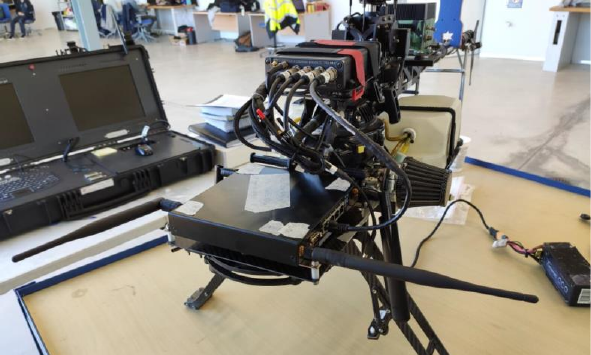}
    \caption{Communications payload installed in the ALPHA~800 UAV.}
    \label{fig2:UAV800}
\end{figure}

\section{System Setup} \label{secIII:System setup}

The system setup of this paper consists of one UAV acting as a Flying Access Point (FAP) and one UE on the ground. The ATLAS site, depicted in~\cref{fig1:hangar} served as the location for conducting the experimental measurements. The UAV is an Alpha 800\footnote{Alpha Unmanned Systems (Spanish company)} and carries a wireless communications module as payload, based on a PC Engines APU4D4 System Board running OpenWRT 19.07.8, and including a Mikrotik R11E-2HPND Wi-Fi interface capable of Multiple Input Multiple Output (MIMO) 2x2 to provide wireless connectivity to the UE. Two omni-directional 2.4~GHz antennas with a gain of 5~dBi are horizontally mounted directly on the communications payload module, which is fixed in the front of the UAV, depicted in ~\cref{fig2:UAV800}. Both antennas are used for the Wi-Fi link to serve the UE, with one pointing its radiation pattern North/South and the other East/West, considering the "North" as the head of the UAV. The UE is a Xiaomi Mi~9T Android smartphone carried by a person at an altitude of approximately 1.3~m above ground. This UE was selected with single antenna capability as the worst-case scenario. If a smartphone with MIMO (2 antennas or more) was used, better network performance would be expected compared to the baseline assessed in this work. Note that this UE is connected to the UAV only through the Wi-Fi link. The IEEE 802.11n (Wi-Fi~4) standard was used, operating in channel 1 with a bandwidth of 20~MHz. A Tx power of 20~dBm and 30~dBm for the UE and UAV were used, respectively. Finally, Minstrel-HT was used as the Wi-Fi MAC auto-rate adaptation mechanism.\looseness=-1

The UAV is connected to the Internet through an LTE link to a local LTE Base Station~(BS). This link is supported by two omni-directional triband antennas that are vertically mounted in the rear of the communications payload. The LTE BS is located 120~m away from the hangar as depicted in~\cref{fig1:hangar}. This BS provides LTE coverage in Band 3~(1.8~GHz) without carrier aggregation, which leads to theoretical throughput values up to 150~Mbit/s for downlink and 50~Mbit/s for uplink. The real throughput measured in the UAV at a distance of 100~m from the LTE BS averaged 114~Mbit/s for downlink and 55~Mbit/s for uplink.\looseness=-1

To generate enough traffic to saturate the Wi-Fi link and assess its performance, we used an \emph{iperf3} TCP server installed on the UAV. The traffic flow was generated from the UAV to the UE. To measure the RSSI at each Wi-Fi antenna on the UAV side, we used \emph{tcpdump}. We also used an \emph{iperf3} client application on the UE to monitor the throughput.\looseness=-1

\section{Field Trial} \label{secIV:FIELD TRIAL}
 The performance of the wireless channel between the UAV and the UE is characterized in terms of the average throughput of a TCP flow sent from the UAV to the UE and also the RSSI (in~dBm) of each received packet measured at each antenna of the UAV, considering different UAV headings and distances.\looseness=-1

Three experimental scenarios were considered and are detailed in what follows:

\subsubsection{Scenario A}

In this experiment, the UAV was positioned 200~m away from the UE (horizontal distance), hovering at an altitude of 50~m Above Ground Level~(AGL). The Euclidean distance between the UE and UAV was 206~m. The UAV was rotated in incremental steps of 45º up to 360º (a full rotation). At each step, the RSSI was measured at the UAV, and the downlink TCP throughput measured at the UE was recorded. This experiment allowed the evaluation of the antennas' ERP and its impact on the measured throughput, since the UAV's body changes this pattern by obstructing and reflecting the signal, depending on the relative heading between the UAV and the UE.\looseness=-1

\subsubsection{Scenario B}
In this experiment, we moved the UAV away from the UE in steps of 25~m, while maintaining the UAV at 50~m AGL, with a heading of 180º relative to the UE. The UAV then came back towards the UE, repeating the same waypoints as before, but with an opposite heading of 0º (UAV head pointing to the UE). At each step, the RSSI measured at UAV and the downlink TCP throughput measured at UE was evaluated.\looseness=-1 

\subsubsection{Scenario C}

In this experiment, the UE was located 1.42 km from the LTE BS, while the UAV was between the UE and the LTE BS (1.2~km away from the LTE BS and 220~m away from the UE). We measured 10 times the Internet throughput achieved when the UE is directly connected to the LTE BS (1.42~km link) compared to when the UE is connected to the LTE BS through the UAV using Wi-Fi (220~m Wi-Fi link + 1.2~km LTE link).\looseness=-1

\section{Experimental Results} \label{secV:EXPERIMENTAL RESULTS}

This section shows the results of the three experimental scenarios presented in Section IV. The results are presented in separate plots for each scenario and performance metric. The RSSI measurements along the distance are also compared with theoretical models, such as Friis and two-ray ground-reflection models.\looseness=-1
\begin{figure}
	\centering
	\includegraphics[width=0.9\linewidth]{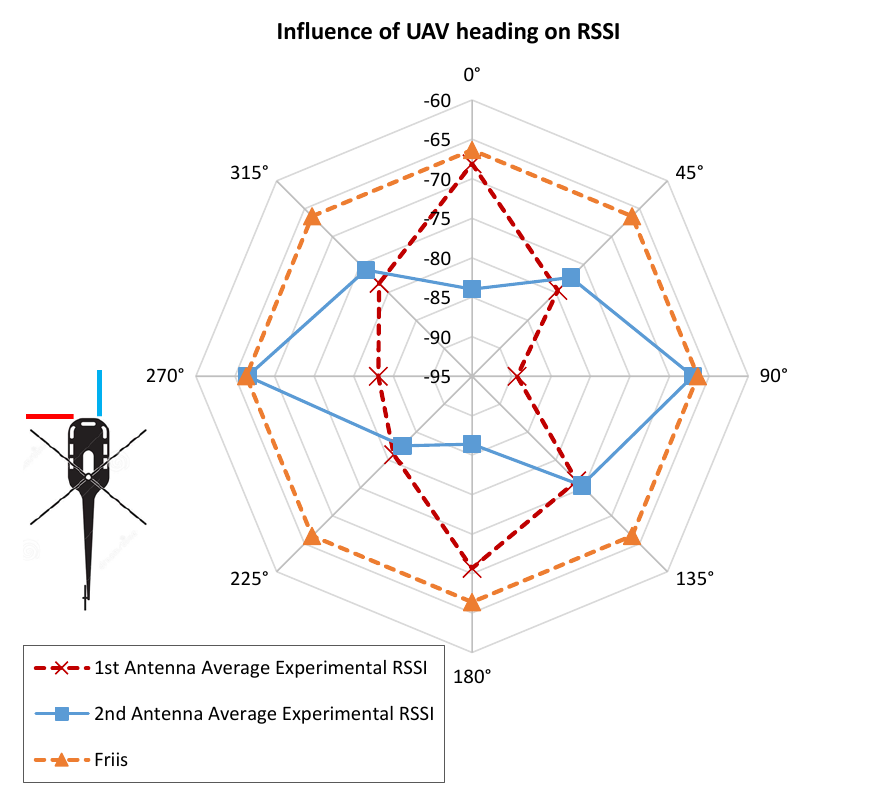}
    \caption{Scenario A -- Measured RSSI of the packets received at both UAV antennas, depending on the relative heading between the UAV and the UE at a Euclidean distance of 206 m, compared to Friis with an isotropic antenna.}
    \label{fig3}
\end{figure}

\subsection{Scenario A: RSSI and Throughput vs. UAV Heading}

\cref{fig3} presents a radar diagram with the measured RSSI of the packets received at the UAV antennas, depending on the relative heading between the UAV and the UE at a Euclidean distance of 206~m. The 1st Antenna (red color) represents the antenna oriented North/South, and 2nd Antenna (blue color) represents the antenna positioned East/West. A simplified representation of a UAV with a red and a blue antenna, depicted in ~\cref{fig3}, was also added to help interpret the radar diagram. The orange line represents the RSSI of the Friis theoretical model, considering an isotropic antenna, and represents the peak RSSI expected for each antenna when aligned with the direction of maximum gain. We can conclude that the radiation pattern is narrow and does not cover the sides of the UAV to provide consistent Wi-Fi coverage to a specific area. Nevertheless, considering both antennas as a whole, which is possible when considering MIMO, the signal of the same stream being received by multiple antennas can be combined.\looseness=-1

\begin{figure}
    \centering
    \subfloat[Effective resulting RSSI of the two antennas for one spatial stream when summing the signal received in both of them at a Euclidean distance of 206 m, compared to Friis with an isotropic antenna.]{
        \includegraphics[width=0.9\linewidth]{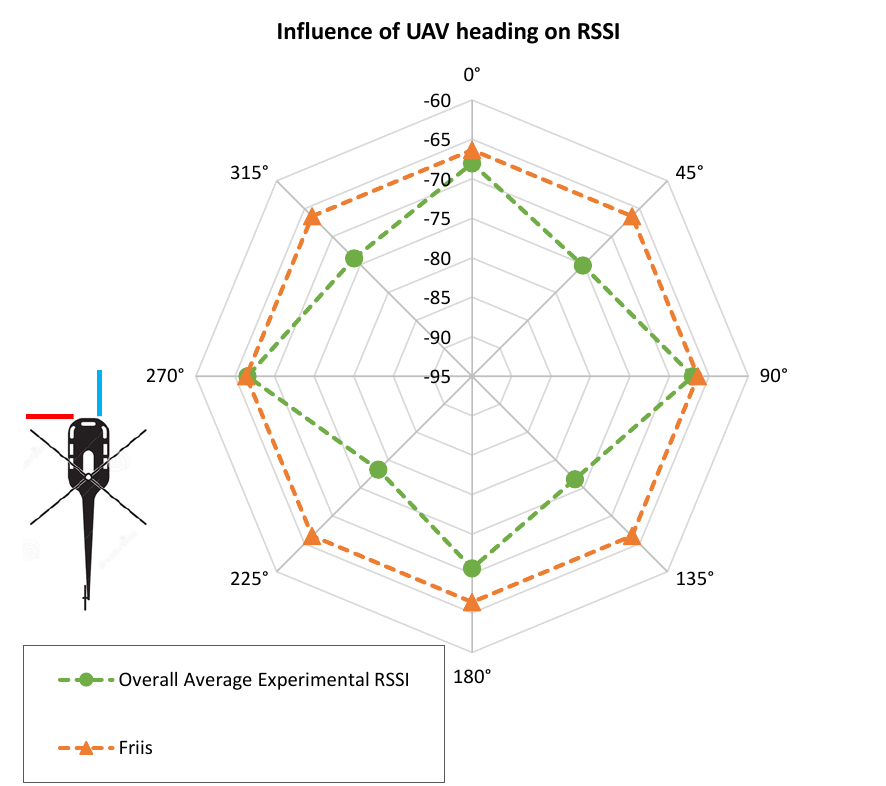}
        \label{fig4a}
    }
    \vfil
    \subfloat[Maximum downlink throughput for the same UAV headings, measured at UE.]{
        \includegraphics[width=0.9\linewidth]{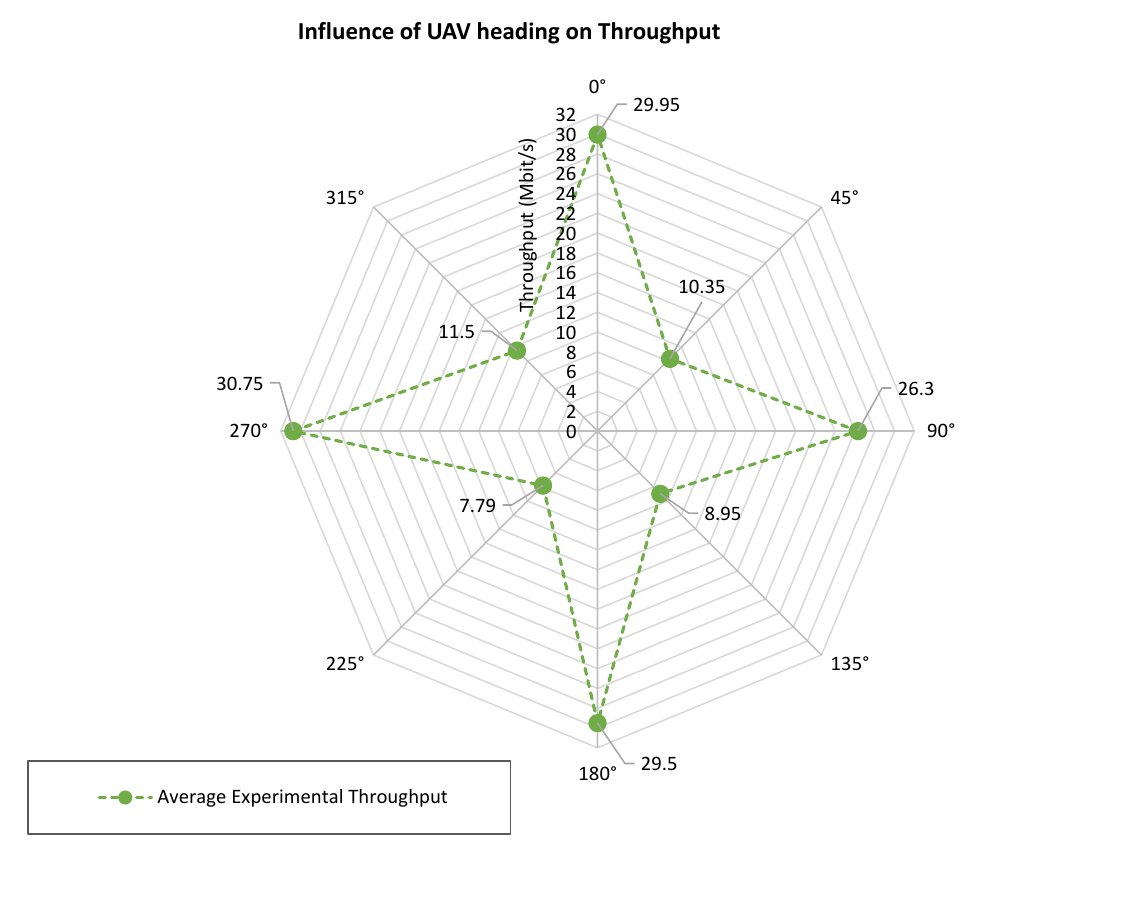}
        \label{fig4b}
    }
    \caption{Scenario A -- Measured RSSI of the antenna system (sum of both antennas), and downlink TCP throughput at the UE, depending on the relative heading between the UAV and the UE.}
    \label{fig4}
\end{figure}

\cref{fig4a} shows the effective resulting RSSI for one spatial stream when summing the power (in dB) of the signals received in both antennas. Note that the radar diagrams in \cref{fig3} and \cref{fig4a} can also be interpreted as ERP diagrams if we consider the orange line as the 0~dBm reference. Therefore, since the antennas are dipoles with omni-directional radiation patterns, and considering their mounting orientations, the resulting ERP met our expectations. Furthermore, it was important to see in effect the capability of the MIMO-capable antenna system to be able to transparently combine the signal of both antennas related to a single stream.\looseness=-1

In~\cref{fig4b}, the maximum downlink throughput for the same UAV headings is depicted. The values presented depict the best and worst-case scenarios, showing that the throughput can vary from approximately 30~Mbit/s (N, S, E, and W) to approximately 10~Mbit/s (NE, SE, SW, and NW). In general, the relationship between the RSSI and the throughput is complex and depends on various factors such as the Signal-to-Noise Ratio~(SNR), Modulation and Coding Scheme~(MCS) being used, and other channel conditions such as interference and occupancy. For example, a strong signal may still experience a high level of interference, which can reduce the throughput. However, in this paper, when comparing~\cref{fig4a} and \cref{fig4b}, we can state that we were achieving the expected values of TCP throughput for the experienced RSSI values due to the low noise floor of -95~dBm and the lack of interference. Additionally, from~\cref{fig4a} we can conclude that at 180º the RSSI is lower than at 0º, which is expected due to the obstruction of the UAV's body.\looseness=-1

\begin{figure}
	\centering
	\includegraphics[width=0.7\linewidth]{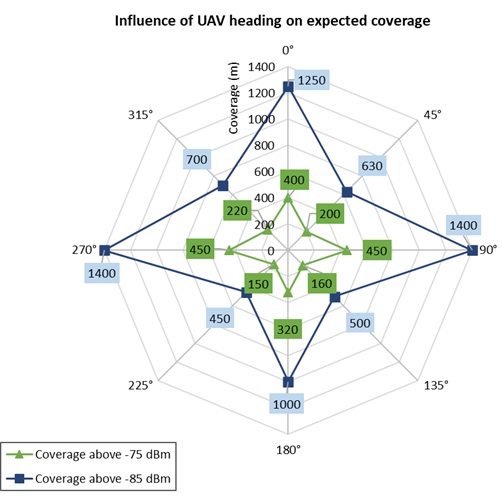}
	\caption{Scenario A -- Resulting expected coverage (in meters) for an RSSI of -85~dBm (link still stable but with low throughput) and -75~dBm (link with still good performance for multiple video streams) cut-offs.}
	\label{fig5}
\end{figure}

\cref{fig5} represents the resulting expected coverage for two RSSI cut-offs: i) a -85~dBm link still stable with a SNR of 10~dB, but with low throughput; and ii) a -75~dBm link with 20~dB of SNR, which still provides good performance for multiple video streams. This considers the measurements at the UAV, which is the worst-case scenario due to the link asymmetry of 10~dB.\looseness=-1

\subsection{RSSI and Throughput vs. Distance}

\begin{figure}[h]
    \centering
    \subfloat[UAV moving away from UE in steps of 25~m, while maintaining a flight at 50~m AGL.]{
        \includegraphics[width=0.9\linewidth]{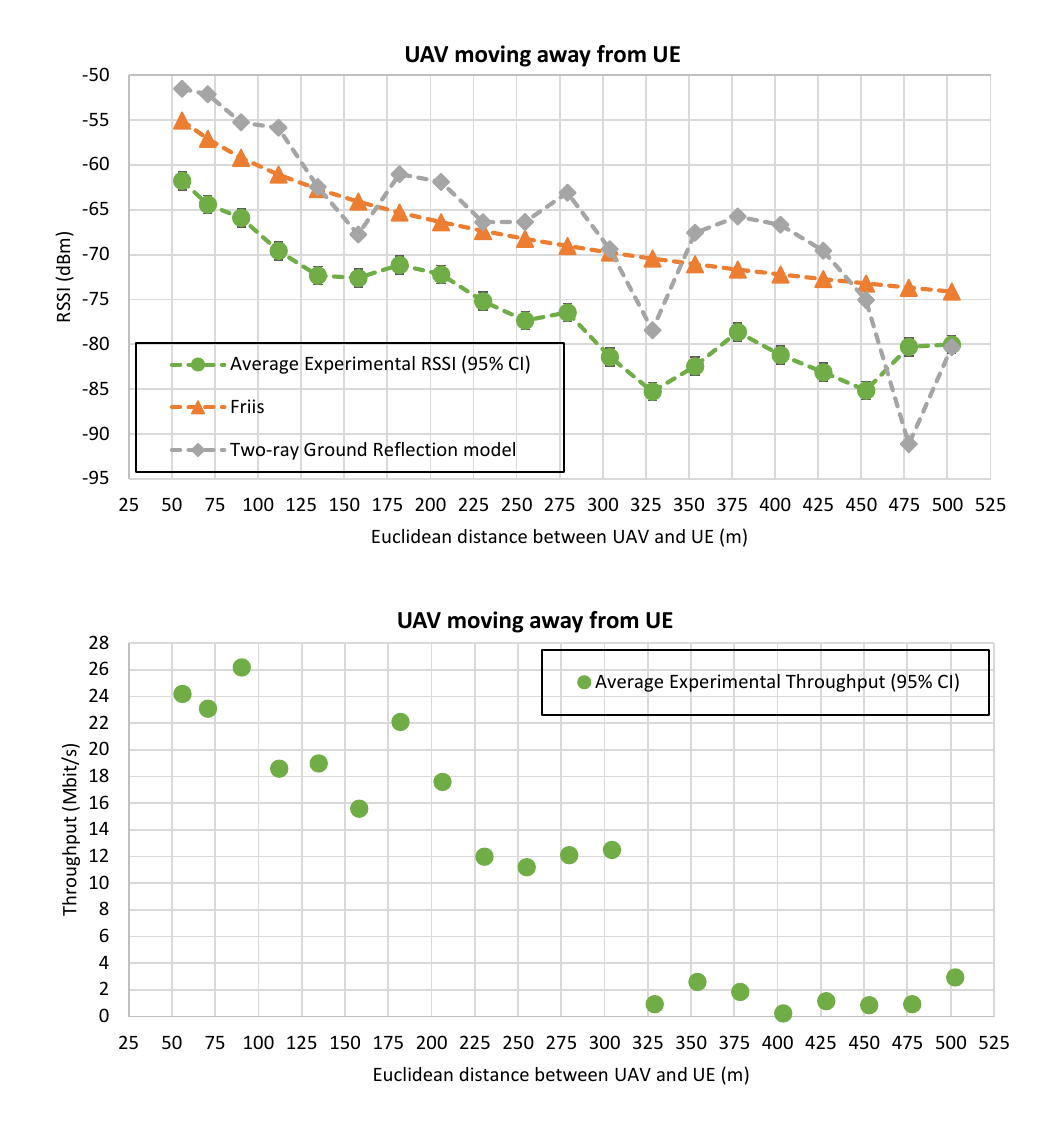}
        \label{fig6a}
    }
    \vfil
    \subfloat[UAV comes back towards the UE, repeating the same waypoints as before.]{
        \includegraphics[width=0.9\linewidth]{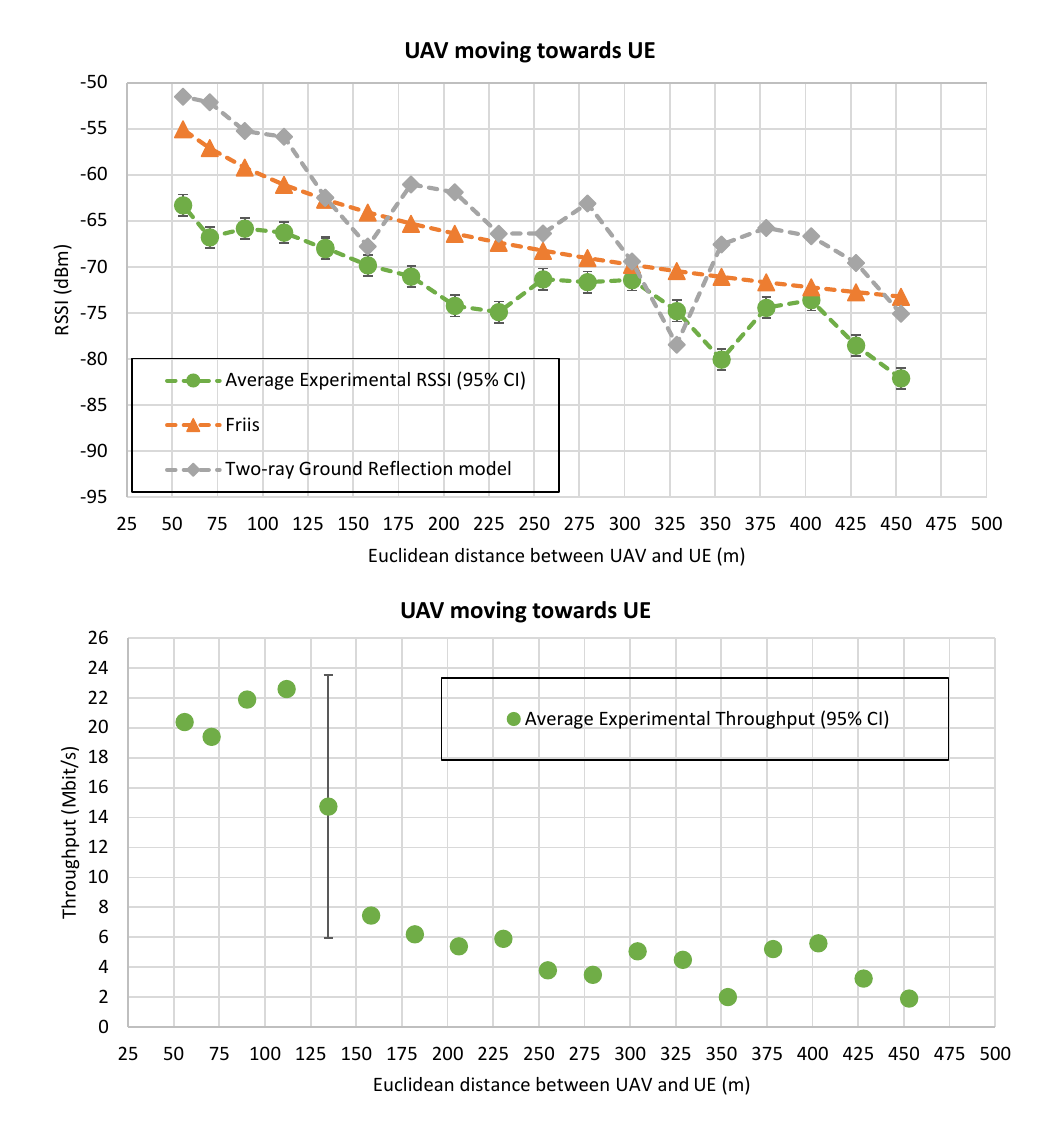}
        \label{fig6b}
    }
    \caption{Scenario B -- Experimental RSSI measured on the UAV compared to the Friis and two-ray ground baselines and the downlink throughput measurements on the UE when the UAV is moving away from the UE and vice-versa.}
    \label{fig6}
\end{figure}

\cref{fig6a} depicts the experimental RSSI results compared to the Friis and two-ray ground reflection propagation loss models, as well as the downlink throughput when the UAV moves away from the UE; \cref{fig6b} shows these parameters when the UAV comes back toward the UE.\looseness=-1

As expected, due to obstruction of the UAV's body, the RSSI was lower when the UAV moved away (180º~heading) than when the UAV was coming back to the UE (0º~heading). The lower throughput values observed during the UAV's return are attributed to a limitation of the Minstrel-HT auto rate mechanism~\cite{6362819}, which is known to delay the increase of the data rate when link conditions improve rapidly. However, upon refueling the UAV at a specific distance and repeating the speed test at the same exact coordinates, the throughput improved significantly as depicted in~\cref{fig6b} at 130~m. In conclusion, it may be necessary to restart the Wi-Fi card to clear the Minstrel-HT history and expedite the process of finding an optimal rate. Furthermore, when the UAV moves away from the UE, a steep decline in throughput occurs, as shown in~\cref{fig6a} between the distances of 300~m and 325~m. This decline is primarily due to the asymmetry of the link, with a 10~dB advantage in favor of the downlink resulting from the differences in Tx power. Specifically, the MAC acknowledgments (ACKs) are sent at 24~Mbit/s whenever data packets are received at 24~Mbit/s or higher using the fast ACK mechanism. Although there is a favorable SNR for higher MCSs in the A2G direction, eventually the G2A SNR becomes too low to support successful acknowledgments at 24~Mbit/s. Consequently, the data packets are "forced" to be sent at lower MCSs to ensure that the ACKs are generated below 24~Mbit/s.\looseness=-1

\begin{figure}
	\centering
	\includegraphics[width=\linewidth, height=5.5cm]{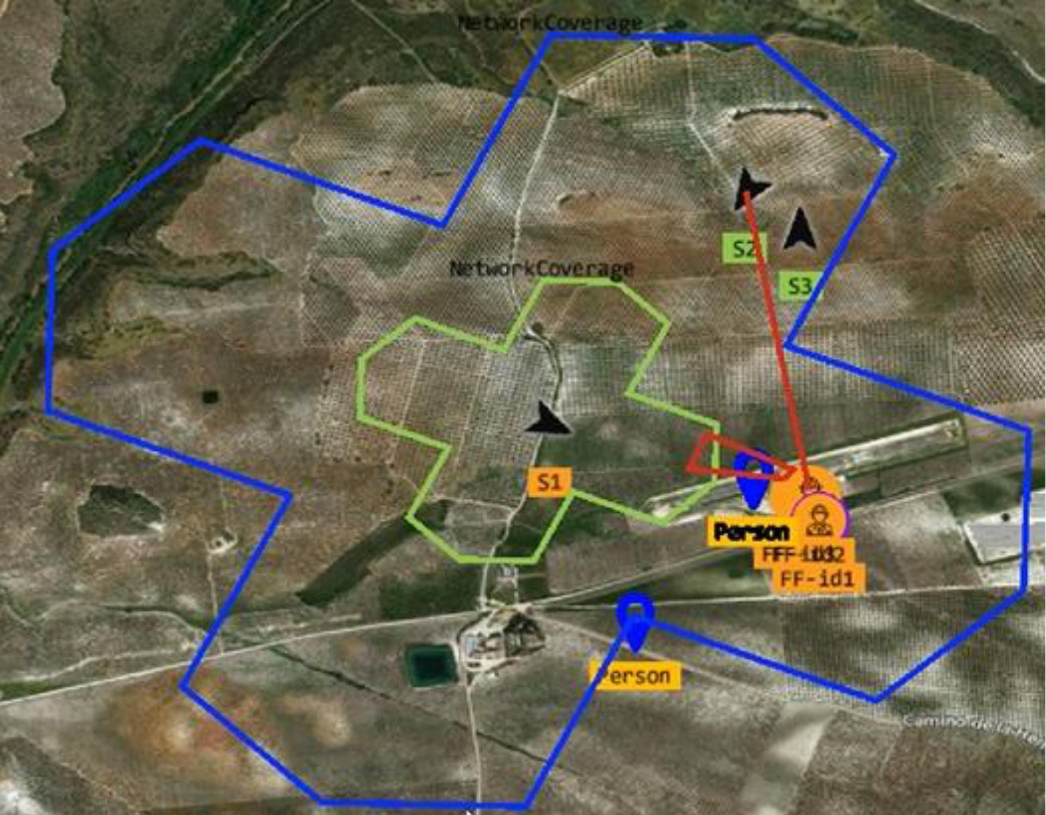}
	\caption{Overlay polygon used in the H2020~ResponDrone project Command Center interface. The blue and green colors represent the same -85~dBm and -75~dBm cut-offs, respectively.}
	\label{fig7}
\end{figure}

\cref{fig7} represents an overlay polygon of the experimental effective Wi-Fi network coverage provided by a UAV (as depicted in ~\cref{fig5}) for a UE with a single antenna, considering its position and heading. The blue and green lines represent, respectively, the same -85~dBm and -75~dBm cut-offs discussed above. For the green area, the throughput was expected to be between 10~Mbit/s and 30~Mbit/s, and the coverage for the N, S, E, and W directions had a range of up to 400~m. Coverage in the NE, SE, SW, and NW directions had a range of up to 200~m. For the blue area, the coverage for the N, S, E, and W directions had a range of up to 1100~m. Coverage in the NE, SE, SW, and NW directions had a range of up to 550~m. For UEs with two antennas, the coverage range is expected to be higher, since we have successfully tested a coverage range of 1500~m, with an RSSI at the UAV side averaging -90~dBm, while being able to successfully browse web pages.\looseness=-1

\subsection{Scenario C: Internet Throughput vs LoS}

\begin{figure}
	\centering
	\includegraphics[width=\linewidth, height=5cm]{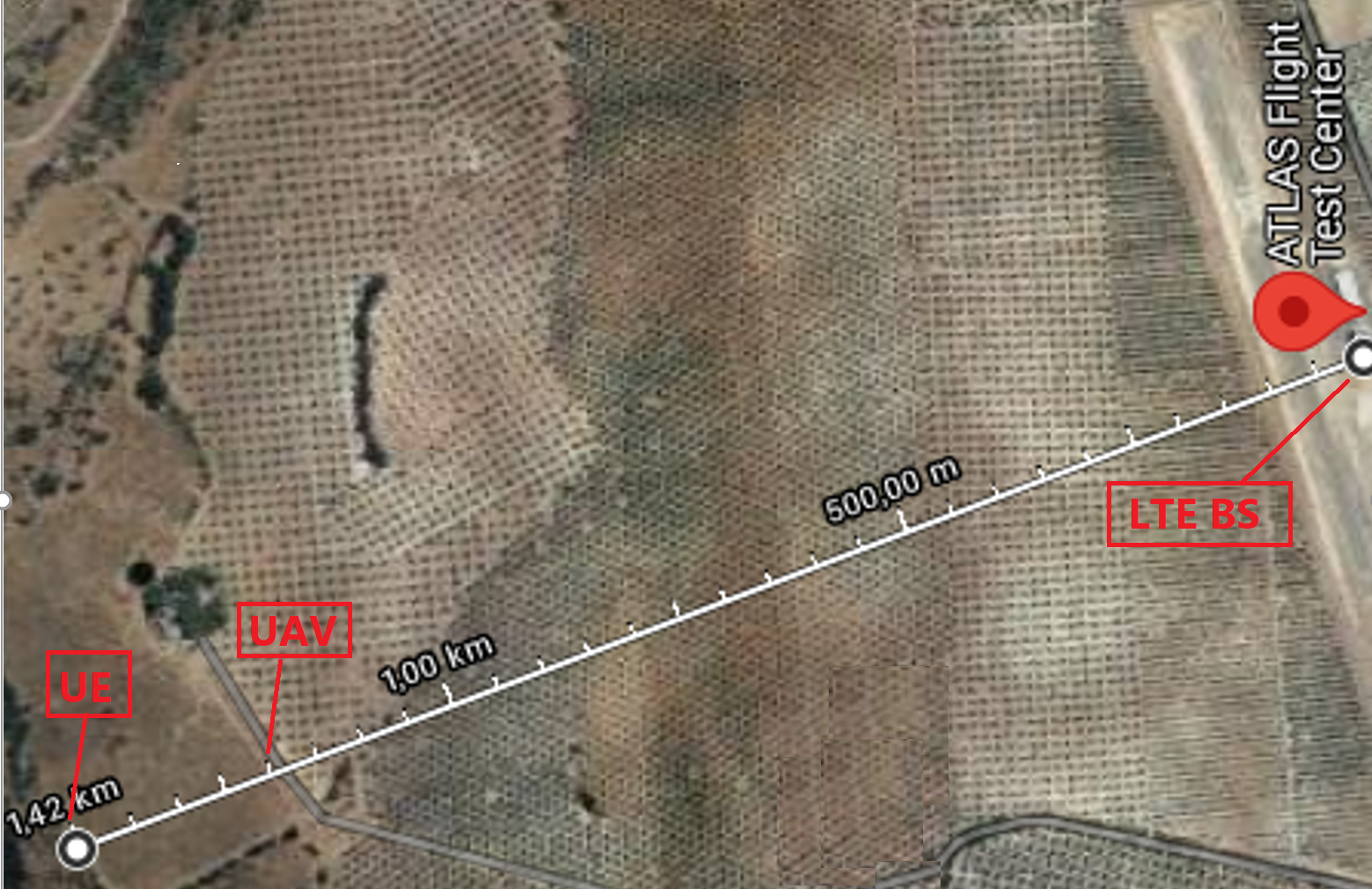}
	\caption{Scenario C -- Distance of the UE to the LTE BS at the hangar. The UE was connected to the Internet either directly via LTE or via Wi-Fi through the UAV, which is then connected to the LTE BS.}
	\label{fig8}
\end{figure}

As depicted in~\cref{fig8}, when the UE was connected directly to the LTE BS, it achieved an average of 13~Mbit/s. However, when the UE was connected to UAV by Wi-Fi, which was relaying traffic to the LTE BS, it reached an average of 21~Mbit/s at the same location, representing a gain of 1.6x. The terrain topography and existing trees blocked the radio LoS between the UE and the LTE BS. However, by going through the UAV, those obstacles were circumvented and radio LoS was ensured.

\section{Conclusions} \label{sec:Conclusions}
This work provided an accurate experimental model that describes the A2G and G2A channels. Our experiments revealed that real-world antenna radiation patterns can be heterogeneous, and the impact of the UAV's body and heading on RSSI should be considered when designing airborne communications systems. As an example, in this specific setup used in the ResponDrone project, the RSSI is generally lower when the UAV is oriented away from the UE compared to when it is pointing towards the UE. We found that UEs with different antenna configurations can affect the connectivity range of the UAV, emphasizing the importance of optimizing antenna design for UAV communications systems. Our findings also showed that Minstrel-HT was able to quickly reduce the MCS being used when the SNR was getting lower (i.e. when the UAV was moving away from the UE), but the opposite was not true since it remained using sub-optimal MCSs in higher SNR link conditions. Clearing the Minstrel-HT link statistics helped it converge faster to an optimal MCS for the observed link SNR. Furthermore, we also concluded that the Fast-ACK mechanism of Wi-Fi was actually degrading the link throughput due to the observed link asymmetry.\looseness=-1

In conclusion, this study offers valuable insights into the potential of UAVs for providing on-demand and dynamic wireless connectivity and the importance of considering UAV heading and antenna configurations in real-world scenarios. Our findings are valuable for future research and development of UAV communication systems and contribute to the optimization of wireless connectivity in various applications.\looseness=-1 

Future works include optimizing antenna designs for UAV communications systems and different UE antenna configurations. Additionally, to enhance rate adaptation mechanisms such as Minstrel-HT to better handle rapidly changing link conditions and address the degradation caused by the Fast-ACK mechanism in asymmetric links.\looseness=-1


\bibliographystyle{IEEEtran}
\bibliography{references}

\end{document}